Letter

# Primordial Dirac leptogenesis

Aqeel Ahmed,[*] Juan P. Garcés,[†] and Manfred Lindner[‡]
*Max-Planck-Institut für Kernphysik, Saupfercheckweg 1, 69117 Heidelberg, Germany*

(Received 7 January 2026; accepted 11 May 2026; published 10 June 2026)

We present a novel realization of Dirac leptogenesis based on the postinflationary reheating phase of the early Universe. An asymmetry generated within the scalar sector via $CP$-violating and out-of-equilibrium inflaton decays is transferred to chiral neutrinos through Yukawa interactions and then to baryons via electroweak sphalerons. We describe in detail a minimal realization of this mechanism that naturally accommodates small neutrino Yukawa couplings and results in contributions to the effective number of relativistic species, $N_{\rm eff}$, testable in upcoming cosmological observations.

*Introduction.* The observed baryon asymmetry of the Universe (BAU) remains one of the most pressing shortcomings of the standard model (SM). Any dynamical mechanism aiming to explain the BAU from symmetric initial conditions must fulfill the three Sakharov conditions [1]: baryon number (B) violation, C and $CP$ violation, and departure from thermal equilibrium. While the SM formally satisfies these criteria, the generated asymmetry is many orders of magnitude too small [2–4], motivating the search for physics beyond the standard model (BSM).

A well-established SM extension that can successfully generate the observed BAU is *Majorana leptogenesis* [5], in which heavy right-handed Majorana neutrinos generate a lepton number (L) asymmetry through their $CP$-violating and out-of-equilibrium decays (see Ref. [6] for a review). Electroweak sphalerons then partially transfer this lepton asymmetry into the baryonic sector, conserving $\mathrm{B-L}$ and violating $\mathrm{B+L}$ [7–9].

In contrast, *Dirac leptogenesis* [10] achieves successful leptogenesis without violating the global lepton number. In this framework, neutrinos are Dirac particles, and consistency with light neutrino mass measurements [11–14] requires the associated Yukawa couplings to be very small, $y_\nu \lesssim \mathcal{O}(10^{-11})$. Provided there is some mechanism that sources an initial asymmetry between left- and right-handed neutrinos and their antiparticles, sphaleron processes partially convert the left-handed neutrino asymmetry into a baryon asymmetry, while the right-handed neutrinos remain out of equilibrium. This allows a nonvanishing BAU to freeze out and persist until today. The theoretical viability and phenomenological implications of Dirac leptogenesis have motivated numerous studies in the past [15–17], as well as more recently [18–22].

Dirac leptogenesis does not depend on a specific mechanism for generating the initial neutrino asymmetry, as long as it occurs above the electroweak scale. In the original realization [10], this asymmetry arises from the out-of-equilibrium decays of two heavy scalar doublets into SM leptons. The smallness of the SM neutrino Yukawa couplings, however, precludes identifying one of these doublets with the SM Higgs, as the resulting asymmetry would be too small to account for the observed BAU.

In this Letter, we propose a new implementation of Dirac leptogenesis in which the required initial asymmetry is generated through early Universe reheating dynamics, with the Higgs doublet playing an important role. The general aspects of the proposed mechanism are summarized in Fig. 1. Starting from $\mathrm{L}_L = \mathrm{L}_R = 0$, with $L_{L/R}$ the net lepton number in left/right-handed neutrinos, there is a first period ⓟ where a chiral lepton asymmetry is produced via lepton number–preserving interactions, i.e., $\Delta\mathrm{L} = \mathrm{L}_L + \mathrm{L}_R = 0$, but $\mathrm{L}_L = -\mathrm{L}_R \neq 0$. This is achieved via $CP$-violating and out-of-equilibrium inflaton decays during reheating, which produce an asymmetric Higgs sector that transfers its asymmetry into neutrinos. The left-handed neutrino asymmetry is then converted to a baryon asymmetry via weak sphaleron processes ⓢ ($\Delta\mathrm{B} = \Delta\mathrm{L}_L$), which become inefficient for temperatures below the electroweak phase transition, $T \lesssim T_{\rm EW} \approx 160$ GeV. Finally, left-right equilibration ⓔ of the neutrino asymmetry occurs at $T \ll T_{\rm EW}$ and, as a result, a nonvanishing final baryon asymmetry, $\Delta\mathrm{B} = \Delta\mathrm{L} \neq 0$, is frozen out.

Unlike the Higgsogenesis framework [23,24], where a Higgs asymmetry directly feeds the baryon asymmetry and requires a strong first-order electroweak phase transition to avoid washout, the scenario proposed here operates efficiently without such an assumption. Furthermore, while baryogenesis mechanisms tied to inflationary dynamics have been considered in Refs. [25–31], the present study provides a distinct realization based on the Dirac leptogenesis mechanism.[1]

[*]Contact author: aqeel.ahmed@mpi-hd.mpg.de
[†]Contact author: juan.garces@mpi-hd.mpg.de
[‡]Contact author: lindner@mpi-hd.mpg.de

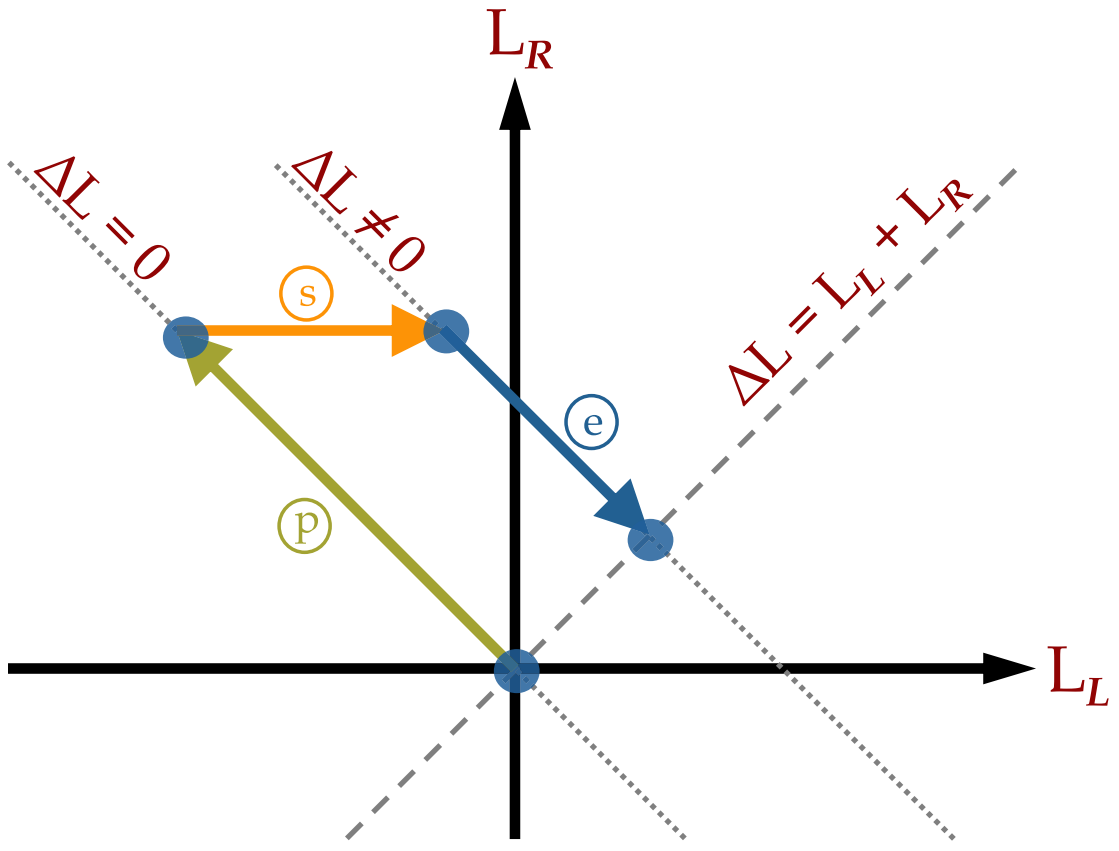

FIG. 1. Schematic illustration of *primordial Dirac leptogenesis*. The inflaton decays produce an initial chiral neutrino asymmetry (ⓟ), which is partially converted into a baryon asymmetry via weak sphalerons (ⓢ). Left-right equilibration of neutrinos (ⓔ) occurs at much lower temperatures, ensuring the freeze-out of the baryon asymmetry.

*A minimal realization.* In the following, we describe a minimal realization of the mechanism mentioned above, although we emphasize that the underlying principle can be readily extended to a variety of other scenarios.

We extend the SM with three BSM fields: the inflaton $\phi$, an inert Higgs doublet $\hat{h}$, and right-handed neutrinos $\nu_R$, which are Dirac partners of the SM left-handed neutrinos $\nu_L$. We charge all of these under a discrete $\mathbb{Z}_2$ symmetry,

$$\phi \to -\phi, \qquad \hat{h} \to -\hat{h}, \qquad \nu_R \to -\nu_R. \tag{1}$$

The $\mathbb{Z}_2$-invariant part of the Lagrangian relevant for the dynamics of interest reads

$$\mathcal{L} \supset -(\hat{y}_\nu \overline{\ell}_L \tau \hat{h}^* \nu_R + \text{H.c.}) - V(h, \hat{h}, \phi), \tag{2}$$

where $\tau = i\sigma_2$, with $\sigma_2$ the second Pauli matrix, $\ell_L = (\nu_L, e_L)^T$ denotes the SM lepton doublet, and $h$ is the SM Higgs doublet. Although the $\mathbb{Z}_2$ symmetry introduced above does not by itself forbid Majorana mass terms, we omit them under the assumption that the global lepton number is conserved.[2] The scalar potential takes the most general renormalizable form,

$$\begin{aligned} V(h,\hat{h},\phi) = {} & m_1^2 h^\dagger h + m_2^2 \hat{h}^\dagger \hat{h} + \lambda_1 (h^\dagger h)^2 \\ & + \lambda_2 (\hat{h}^\dagger \hat{h})^2 + \lambda_3 (h^\dagger h)(\hat{h}^\dagger \hat{h}) + \lambda_4 (h^\dagger \hat{h})(\hat{h}^\dagger h) \\ & + \frac{1}{2}\lambda_5 [(h^\dagger \hat{h})^2 + (\hat{h}^\dagger h)^2] + V_{\text{inf}}(\phi) + g_\phi M_{\text{Pl}} \phi h^\dagger \hat{h} \\ & + g_\phi^* M_{\text{Pl}} \phi \hat{h}^\dagger h + \lambda_{\phi h} \phi^2 h^\dagger h + \lambda_{\phi \hat{h}} \phi^2 \hat{h}^\dagger \hat{h}. \end{aligned} \tag{3}$$

The $\phi$-independent terms above reproduce the standard inert doublet potential [38,39], while the rest correspond to the inflaton potential, $V_{\text{inf}}(\phi)$, and its portal interactions with the Higgs doublets. All couplings are taken to be real except for the trilinear inflaton coupling $g_\phi$, whose complex phase provides the sole source of $CP$ violation in this framework.[3] The precise form of the inflaton potential, which governs the inflationary background dynamics, is not essential for the baryogenesis mechanism described below, and is therefore left unspecified.

The relevant inflaton decay diagrams that can generate asymmetric final states during reheating are

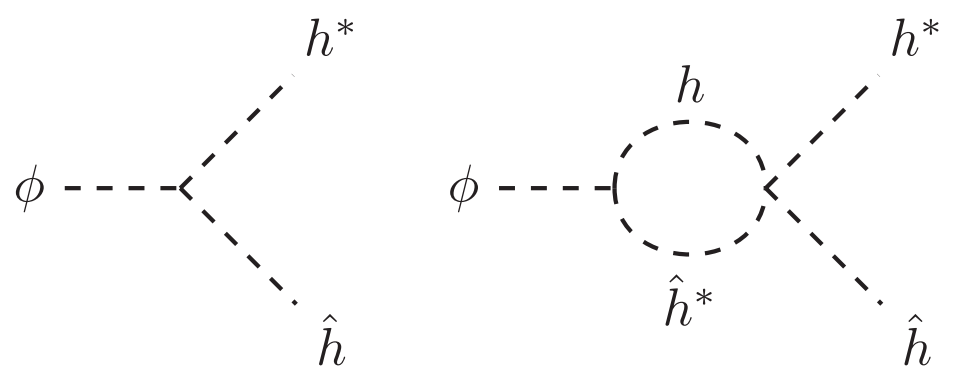

+ c.c. (4)

The interference between these amplitudes induces a scalar asymmetry between $h/\hat{h}^*$ and $h^*/\hat{h}$, provided that the doublets are produced on-shell and that the $CP$-violating phase of $g_\phi$ is nonvanishing. Since inflaton decays occur in an expanding Universe, the out-of-equilibrium condition is naturally fulfilled.

The resulting Higgs asymmetry is subsequently transferred to left- and right-handed leptons and quarks through the Yukawa interactions of the two Higgs doublets. The sphaleron processes ensure that the last Sakharov condition, namely, baryon number violation, is also satisfied, and if the neutrino asymmetry is generated above the sphaleron freeze-out temperature $T_{\text{fo}}^{\text{sph}} \sim T_{\text{EW}}$, it will render a nonvanishing final BAU.

[1] We refer to [19] and [32] for an implementation of Dirac leptogenesis in the context of the Affleck-Dine [33] and Froggatt-Nielsen [34] mechanisms, respectively.

[2] Since lepton number is imposed as a global symmetry, Planck scale–suppressed operators can give rise to lepton number–violating operators, such as small Majorana masses $m_N$ for the right-handed neutrinos. However, their impact on relevant processes is suppressed by $\mathcal{O}(m_N^2/T^2)$ [35–37], and since the relevant temperatures in Dirac leptogenesis satisfy $T \gtrsim T_{\text{EW}}$, these effects are expected to be negligible.

[3] Although $\lambda_5$ could, in principle, be complex, an appropriate phase redefinition of the Higgs doublets renders it real without loss of generality.

In the following, we neglect higher-order radiative corrections induced by inflaton-Higgs couplings and assume that the inflaton potential, $V_{\rm inf}(\phi)$, solely governs the background dynamics. We can then write the coherently oscillating inflaton field as

$$\phi(t) = \varphi(t) \sum_{k=-\infty}^{\infty} \mathcal{P}_k \, e^{ik\omega t}, \tag{5}$$

where $\mathcal{P}_k$ are Fourier coefficients of the oscillation mode with frequency $\omega$, and $\varphi(t)$ is a slowly varying envelope function (see, e.g., Refs. [40–42] for further details). For a quadratic inflaton potential, $\mathcal{P}_1 = 1/2$ and $\omega = m_\phi$ are constant, while more general potentials excite multiple harmonics with $\mathcal{P}_k \sim \mathcal{O}(1)$ and a time-dependent $\omega \sim m_\phi$ during reheating.

The $CP$ asymmetry between the conjugate decay channels $\phi \to h^* \hat{h}$ and $\phi \to h \hat{h}^*$ arises from the interference between the tree- and absorptive loop-level amplitudes. It is given by

$$\begin{aligned} \Delta\Gamma_\phi &\equiv \Gamma(\phi \to h^* \hat{h}) - \Gamma(\phi \to h \, \hat{h}^*) \\ &\approx \frac{3\, M_{\rm Pl}^2 \, \varphi(t)^2}{64\, \pi^2 \, \rho_\phi} |g_\phi|^2 \, \lambda_5 \, \sin(2\theta) \sum_{k=-\infty}^{\infty} |\mathcal{P}_k|^2 \, k\, \omega, \end{aligned} \tag{6}$$

in the regime where $k\omega \gg m_1 + m_2$,[4] with $\Gamma(i \to j)$ denoting the interaction rate of the process $i \to j$, $\theta \equiv \arg(g_\phi)$ the irremovable $CP$-violating phase, and $\rho_\phi$ the inflaton energy density. Possible additional $CP$-violating contributions from inflaton scattering processes involving quartic interactions are neglected, as they are expected to be subdominant.

The total inflaton decay width $\Gamma_\phi$, which governs the reheating dynamics, is dominated by tree-level processes and reads

$$\Gamma_\phi \approx \frac{M_{\rm Pl}^2 \, \varphi(t)^2}{8\, \pi \, \rho_\phi} |g_\phi|^2 \sum_{k=-\infty}^{\infty} |\mathcal{P}_k|^2 \, k\, \omega. \tag{7}$$

Combining Eqs. (6) and (7) we obtain the Higgs sector $CP$ asymmetry

$$\epsilon \equiv \frac{\Delta\Gamma_\phi}{2\Gamma_\phi} \approx \frac{3}{16\pi} \lambda_5 \sin(2\theta), \tag{8}$$

which depends solely on the quartic coupling $\lambda_5$ and the irremovable $CP$-violating phase $\theta$, and vanishes if any of these is zero, as expected.

[4]The assumption that the inflaton energy in the decaying mode satisfies $k\omega \gg m_1 + m_2$ can be relaxed with $\mathcal{O}(1)$ corrections, however, the condition $k\omega > m_1 + m_2$ is required to generate a nonvanishing absorptive part of the loop amplitude.

The inert doublet $\hat{h}$ transfers its asymmetry to left- and right-handed neutrinos via decays mediated by the Yukawa interaction of Eq. (2). To maintain consistency with neutrino data, $\hat{h}$ must not acquire a vacuum expectation value (VEV).[5] This is possible if

$$\lambda_2 > 0, \quad \lambda_3 > -\frac{m_2^2}{v^2}, \quad \lambda_{345} > -\frac{m_2^2}{v^2}, \quad \bar{\lambda}_{345} > -\frac{m_2^2}{v^2}, \tag{9}$$

where $v = \langle h \rangle$ is the SM Higgs VEV, $\lambda_{345} \equiv \lambda_3 + \lambda_4 + \lambda_5$, $\bar{\lambda}_{345} \equiv \lambda_3 + \lambda_4 - \lambda_5$, and provided that the total potential is bounded from below. Furthermore, in order for the generated scalar asymmetry not to be washed out before the inert doublet decays into neutrinos, we must require that the interactions converting $h \leftrightarrow \hat{h}^*$ and $h^* \leftrightarrow \hat{h}$ remain out of equilibrium until the time of $\hat{h}$ decay, i.e., we must ensure that

$$\Gamma(hh \leftrightarrow \hat{h}^* \hat{h}^*)|_{T_d} < H(T_d), \tag{10}$$

with $T_d$ the temperature at the time of $\hat{h}$ decay, and $H$ the Hubble rate. As mentioned above, we also require $\hat{h}$ to decay before the electroweak phase transition, namely,

$$T_d \gtrsim T_{\rm EW}. \tag{11}$$

The washout processes mentioned above are given by 2-to-2 scatterings mediated by $\lambda_5$. Thus, Eq. (10) gives an upper bound on $\lambda_5$ [24],

$$|\lambda_5| \lesssim 10^{-7} \sqrt{\frac{T_d}{T_{\rm EW}}}, \tag{12}$$

which is important for the present study given that the generated BAU, as will be shown below, is proportional to $\lambda_5$.

We note that the mechanism responsible for generating the Higgs asymmetry is largely independent of the detailed form of the inflationary model. However, we assume that the reheating temperature is greater than the inert Higgs mass, $m_{\hat{h}}$, and the electroweak phase transition temperature, $T_{\rm EW}$. This condition leads to a lower bound on the $CP$-violating inflaton-Higgs coupling $g_\phi$, whose precise value depends on the specific inflationary or reheating potential. Moreover, demanding that inflation is driven by the inflaton potential, $V_{\rm inf}(\phi)$, places an upper bound on $g_\phi$, ensuring that the trilinear inflaton-Higgs interaction remains subdominant with respect to $V_{\rm inf}$ (see, e.g., Ref. [42]).

Finally, if interactions are to respect the $\mathcal{Z}_2$ symmetry of Eq. (1) and we require $\langle \hat{h} \rangle = 0$, the neutrinos remain

[5]Strictly speaking, $\hat{h}$ could have a small vacuum expectation value, but we ignore this possibility for simplicity.

massless at tree level. One way of generating small Dirac neutrino masses is via the SM Yukawa interaction

$$\mathcal{L} \supset -y_\nu \bar{L}\, \tau\, h^* \nu_R + \text{H.c.}, \tag{13}$$

where $y_\nu = m_\nu/\langle h \rangle \simeq 5.7 \times 10^{-12} (m_\nu/\text{eV})$ is naturally small given that it is the only term that explicitly breaks the $\mathbb{Z}_2$ symmetry.[6]

*From scalar to baryon asymmetry.* The scalar asymmetries in the SM Higgs and inert Higgs are communicated to SM fermions and right-handed neutrinos as an asymmetry between chiral states via lepton number–conserving Yukawa couplings, leaving $\text{B} - \text{L}$ unchanged. Assuming symmetric initial conditions ($\text{B} - \text{L} = 0$), sphalerons together with left-right equilibration processes erase any previously generated asymmetry among charged fermions. For neutrinos, however, these processes are never simultaneously efficient because the Yukawa couplings are extremely small, so a chiral asymmetry in the neutrino sector persists. Therefore, the only relevant reservoir of left-handed lepton number, which the sphalerons partially convert to baryons, is that of neutrinos, $\text{L}_L^\nu$. In particular, the surviving baryon number asymmetry $\text{B}_0$ satisfies the relation

$$\text{B}_0 - \text{L}_L^\nu = \text{L}_R^\nu. \tag{14}$$

The chemical equilibrium equations then give the following relation between the final baryon number asymmetry and the lepton number asymmetry in right-handed neutrinos [43]:

$$\text{B}_0 = -\frac{28}{79} \text{L}_R^\nu. \tag{15}$$

Next, we estimate the baryon asymmetry yield, defined as

$$Y_\text{B} \equiv \frac{n_\text{B} - n_{\bar{\text{B}}}}{s}, \tag{16}$$

with $s$ the entropy density and $n_\text{B}$ ($n_{\bar{\text{B}}}$) the baryon (antibaryon) number densities. For simplicity, we focus on the *drift-and-decay* limit [44–46], where inverse decays and scattering processes can be ignored, and the dynamics of leptogenesis is fully determined by the decay processes alone. This limit is applicable for $T = T_d \sim m_{\hat{h}}$ when [47]

$$\Gamma_{\hat{h}} \ll H(m_{\hat{h}}), \tag{17}$$

with $\Gamma_{\hat{h}}$ the total decay rate of $\hat{h}$,

$$\Gamma_{\hat{h}} \simeq \frac{|\hat{y}_\nu|^2}{16\pi} m_{\hat{h}}. \tag{18}$$

In terms of masses and couplings, Eq. (17) reads

$$|\hat{y}_\nu| \ll 10^{-8} \sqrt{\frac{m_{\hat{h}}}{\text{GeV}}}, \tag{19}$$

i.e., the drift-and-decay limit is applicable for small Yukawa couplings $\hat{y}_\nu$ and/or large masses $m_{\hat{h}}$.

Since right-handed neutrinos are dominantly produced by decays of $\hat{h}$ and $\hat{h}^*$, we can compute the final baryon asymmetry directly from their number densities at the time of decay, namely,

$$Y_\text{B} = -\frac{28}{79} Y_{\Delta\nu_R} \simeq \frac{28}{79} Y_{\Delta\hat{h}}(T_d), \tag{20}$$

with $Y_{\Delta\nu_R} \equiv (n_{\nu_R} - n_{\bar{\nu}_R})/s$ and $Y_{\Delta\hat{h}} \equiv (n_{\hat{h}} - n_{\hat{h}^*})/s$. The asymmetric Higgs yield can be rewritten as

$$Y_{\Delta\hat{h}} = \frac{\Delta\rho_{\hat{h}}}{\langle E \rangle s} = \frac{45\zeta(3)}{2\pi^4} \frac{g_\star}{g_{\star s}} \frac{\Delta\rho_{\hat{h}}}{\rho_R}, \tag{21}$$

where $\Delta\rho_{\hat{h}} \equiv \rho_{\hat{h}} - \rho_{\hat{h}^*}$ is the asymmetric Higgs energy density, $\rho_R$ is the total radiation energy density, and $\langle E \rangle$ is the mean energy of relativistic bosons. Above, $g_\star$ and $g_{\star s}$ denote the effective number of relativistic degrees of freedom contributing to the energy density and entropy density, respectively.

As the Higgs asymmetry is generated during the reheating phase and the scaling for both $\Delta\rho_{\hat{h}}$ and $\rho_R$ remain the same during and after the end of this period, the temperature dependence of $Y_{\Delta\hat{h}}(T)$ is only through $g_\star(T)/g_{\star s}(T)$, which is negligible. The simplified Boltzmann equations for $\Delta\rho_{\hat{h}}$ and $\rho_R$, during the reheating phase, are given by[7]

$$\begin{aligned} \frac{d\Delta\rho_{\hat{h}}}{dt} + 4\,H\,\Delta\rho_{\hat{h}} &\simeq 2\,\epsilon\,\Gamma_\phi\,\rho_\phi, \\ \frac{d\rho_R}{dt} + 4\,H\,\rho_R &\simeq \Gamma_\phi\,\rho_\phi. \end{aligned} \tag{22}$$

Since these equations depend on the evolution of the inflaton energy density during reheating, the result will be sensitive to the shape of the inflaton potential around the minimum. For an inflationary potential that behaves quadratically at field values close to the minimum, one obtains (see Refs. [40–42,47] for further details)

[6]Equation (13) can radiatively generate additional $Z_2$-breaking operators. However, these contributions are proportional to $y_\nu$ and are, therefore, phenomenologically irrelevant.

[7]In standard perturbative reheating scenarios, such as the one considered here, the inflaton decays out of equilibrium, and inverse decays are strongly suppressed.

$$\Delta\rho_{\hat{h}}^{\rm rh} \simeq \frac{8}{5}\epsilon\,\tilde{\Gamma}_\phi\sqrt{3\rho_\phi(a_e)}\,M_{\rm Pl}\,e^{-3N_{\rm rh}/2},$$
$$\rho_R^{\rm rh} \simeq \frac{4}{5}\tilde{\Gamma}_\phi\sqrt{3\rho_\phi(a_e)}M_{\rm Pl}\,e^{-3N_{\rm rh}/2}, \qquad (23)$$

with the superscript "rh" denoting the time at the end of reheating, $N_{\rm rh}$ the number of e-folds during reheating, $\rho_\phi(a_e)$ the inflaton energy density at the end of inflation, and $\tilde{\Gamma}_\phi$ the part of Eq. (7) that is independent of the number of e-folds during reheating. From the above solutions, we obtain[8]

$$\frac{\Delta\rho_{\hat{h}}(T_d)}{\rho_R(T_d)} = \frac{\Delta\rho_{\hat{h}}^{\rm rh}}{\rho_R^{\rm rh}} \simeq 2\epsilon, \qquad (24)$$

and replacing this in Eq. (20) together with Eq. (8) gives the predicted value for the baryon asymmetry yield,[9]

$$Y_{\rm B} \simeq (1.2\times10^{-2})\lambda_5\sin(2\theta). \qquad (25)$$

Interestingly, all of the information about the inflationary potential dictating the reheating dynamics drops out of the ratio of energy densities needed to compute the baryon asymmetry yield.

The measured value of $Y_{\rm B}$ is given by [14]

$$Y_{\rm B}^{\rm obs} \simeq 8.7\times10^{-11}, \qquad (26)$$

which can be achieved in the present scenario for

$$\lambda_5 \simeq 7.4\times10^{-9}\left(\frac{1}{\sin(2\theta)}\right). \qquad (27)$$

Plugging this result into Eq. (12) we obtain a bound for the decoupling temperature of $\hat{h}$,

$$T_d \gtrsim \frac{5\times10^{-3}}{\sin^2(2\theta)}T_{\rm EW}, \qquad (28)$$

which needs to be satisfied in order to protect the scalar asymmetry from washing out. For $\sin(2\theta) \gtrsim \mathcal{O}(0.1)$, this bound is weaker than that of Eq. (11). Requiring that $\hat{h}$ decays before the electroweak phase transition implies

$$\Gamma_{\hat{h}}(T_{\rm EW}) > H(T_{\rm EW}), \qquad (29)$$

which translates to

$$|\hat{y}_\nu| \gtrsim 6\times10^{-8}\sqrt{\frac{T_{\rm EW}}{m_{\hat{h}}}}. \qquad (30)$$

[8]Subleading corrections depending on the inflaton potential exist but are neglected here for simplicity.

[9]For usual reheating temperatures $g_* = g_{*s}$.

Combining the conditions of Eqs. (17) and (29) we conclude that in the drift-and-decay limit, the inert doublet must be significantly heavier that the electroweak scale in order to allow for successful baryogenesis.

*Testability.* The presence of light, relativistic species beyond the standard model increases the total radiation energy density of the Universe. This effect is conventionally parametrized by the *effective number of neutrino species*, $N_{\rm eff}$, defined through

$$\rho_{\rm R} = \rho_\gamma\left[1+\frac{7}{8}\left(\frac{4}{11}\right)^{4/3}N_{\rm eff}\right], \qquad (31)$$

where $\rho_\gamma$ is the photon energy density. In the standard model, neutrino decoupling leads to $N_{\rm eff}^{\rm SM} \simeq 3.045$. Any additional relativistic degrees of freedom contribute to the deviation $\Delta N_{\rm eff} \equiv N_{\rm eff} - N_{\rm eff}^{\rm SM}$. The current bound on this quantity reads [48]

$$\Delta N_{\rm eff} < 0.285 \quad (\text{at } 2\sigma \text{ C.L.}), \qquad (32)$$

while the expected future reach consists of an improvement of more than 1 order of magnitude on this bound [49,50].

All scenarios based on Dirac leptogenesis lead to nonvanishing contributions to $\Delta N_{\rm eff}$ due to the presence of light right-handed neutrinos.[10] General impacts on $\Delta N_{\rm eff}$ due to freeze-in production of these light degrees of freedom is comprehensively discussed in Ref. [51], where the contribution due to decays of a thermalized boson, in our case $\hat{h}$, is estimated to be

$$\Delta N_{\rm eff} \sim 0.4\times\left(\frac{100}{g_*}\right)^{11/6}\left(\frac{T_{\rm EW}}{m_{\hat{h}}}\right)\left|\frac{\hat{y}_\nu}{10^{-7}}\right|^2. \qquad (33)$$

For instance, if $m_{\hat{h}} = 500$ GeV and $\hat{y}_\nu = 10^{-7}$, one expects $\Delta N_{\rm eff} \sim 0.1$, which is in reach of future experiments.[11]

Combining Eqs. (30) and (33), the current bound on $\Delta N_{\rm eff}$ given in Eq. (32) can be recast as a lower limit on the inert doublet mass,

$$m_{\hat{h}} \gtrsim T_{\rm EW}, \qquad (34)$$

which lies within the reach of collider experiments [52].

Moreover, the current nonobservation of neutrinoless double beta decay does not demonstrate that neutrinos are

[10]We refer the reader to Ref. [18] for generic Dirac leptogenesis scenarios and their impacts on $\Delta N_{\rm eff}$.

[11]Note that in the present scenario, $\hat{h}$ is produced nonthermally through inflaton decays. However, gauge interactions bring $\hat{h}$ into kinetic equilibrium with the SM bath well before right-handed neutrinos are produced via freeze-in. Thus, the abundance of $\hat{h}$ during neutrino production is effectively determined by the temperature of the thermal bath and Eq. (33) applies.

Dirac fermions, but it increasingly motivates theoretical frameworks in which the lepton number is conserved, such as Dirac neutrino models (see, e.g., Ref. [53] for a review).

*Summary and conclusions.* We have presented a novel realization of the Dirac leptogenesis framework in which the required neutrino asymmetry originates from the presence of $CP$-violating interactions in the primordial Universe. We emphasize that, unlike previous Dirac leptogenesis models, e.g., based on the out-of-equilibrium decays of heavy doublets [10], our setup generates the primordial asymmetry within the scalar sector as a consequence of the postinflationary reheating dynamics. This provides a dynamical link between inflation, reheating, and the origin of the baryon asymmetry of the Universe.

We have analyzed a minimal scenario where the scalar sector of the standard model is extended by an additional $SU(2)_L$ doublet coupled to both the Higgs and the inflaton, along with right-handed neutrinos that break a dark sector $\mathbb{Z}_2$ symmetry and generate small Dirac neutrino masses. Then, $CP$-violating inflaton decays into the Higgs doublets result in a Higgs asymmetry that is subsequently transferred to neutrinos via Yukawa interactions, and partially converted into a final baryon asymmetry through weak sphaleron processes. We emphasize that, as in Ref. [10], the global lepton number in this scenario remains conserved, which is a major difference from standard Majorana leptogenesis [5]. However, the two scalar doublets present here take very different roles compared to the two heavy doublets in the original Dirac leptogenesis setup [10], mainly because we identify one of them with the standard model Higgs. Because of the tiny neutrino Yukawas, the self-energy and vertex loop contributions discussed in Ref. [10] are strongly suppressed. Consequently, in the minimal scenario described in this work, the $CP$-asymmetric inflaton decays during reheating are the essential origin of the observed baryon asymmetry of the Universe.

Approximate analytic results derived for the reheating and radiation-dominated eras show that a nonzero baryon asymmetry can be obtained in the presence of an irreducible $CP$-violating phase and a quartic interaction between the Higgs doublets. We show that there is available parameter space that can generate the observed baryon asymmetry of the Universe while satisfying the two crucial model constraints: (i) that the freeze-in production of right-handed neutrinos due to $\hat{h}$ decays happens before sphaleron freeze-out, and (ii) that the scalar asymmetry is not washed out before $\hat{h}$ decays.

In summary, we presented a mechanism that can successfully explain the observed baryon asymmetry of the Universe based on a primordial asymmetry generated via $CP$-violating reheating dynamics and the sequestering mechanism of Dirac leptogenesis. It offers a general and testable framework that can be extended to address other open questions of the standard model, including the origin of dark matter [54].

*Acknowledgments.* We thank Julian Heeck and Werner Rodejohann for helpful comments on the manuscript. J. P. G. acknowledges funding from the International Max Planck Research School for Precision Tests of Fundamental Symmetries (IMPRS-PTFS).

*Data availability.* The data that support the findings of this article are not publicly available. The data are available from the authors upon reasonable request.